\documentclass[prl,twocolumn,showpacs,superscriptaddress]{revtex4}

\usepackage{graphicx,verbatim}

\begin{document}

\title{Valence band electronic structure of V$_2$O$_3$: identification of V and O bands}

\author{E. Papalazarou}
\email{evangelos.papalazarou@lasim.univ-lyon1.fr}
\affiliation{Laboratoire d'Optique Appliqu\'{e}e, ENSTA-Ecole Polytechnique, 
91761 Palaiseau, France }
\affiliation{Laboratoire de Spectrom\'{e}trie Ionique et Mol\'{e}culaire, Universit\'{e} Claude Bernard Lyon 1, 69622 Villeurbanne cedex, France}

\author{Matteo Gatti}
\affiliation{Laboratoire des Solides Irradi\'{e}s, Ecole Polytechnique,
CNRS, CEA/DSM, 91128 Palaiseau, France}
\affiliation{European Theoretical Spectroscopy Facility (ETSF) }

\author{M. Marsi}
\affiliation{Laboratoire de Physique des Solides, CNRS-UMR 8502, Universit\'{e}
Paris-Sud, 91405 Orsay, France }

\author{V. Brouet}
\affiliation{Laboratoire de Physique des Solides, CNRS-UMR 8502, Universit\'{e}
Paris-Sud, 91405 Orsay, France }

\author{F. Iori}
\affiliation{Laboratoire des Solides Irradi\'{e}s, Ecole Polytechnique,
CNRS, CEA/DSM, 91128 Palaiseau, France}
\affiliation{European Theoretical Spectroscopy Facility (ETSF) }

\author{Lucia~Reining}
\affiliation{Laboratoire des Solides Irradi\'{e}s, Ecole Polytechnique,
CNRS, CEA/DSM, 91128 Palaiseau, France}
\affiliation{European Theoretical Spectroscopy Facility (ETSF) }

\author{E. Annese}
\affiliation{Laboratorio Nazionale TASC-INFM-CNR, in Area
Science Park, I-34012 Trieste, Italy}

\author{I. Vobornik}
\affiliation{Laboratorio Nazionale TASC-INFM-CNR, in Area
Science Park, I-34012 Trieste, Italy}

\author{F. Offi}
\affiliation{CNISM and Dipartimento di Fisica Universit\`{a} Roma Tre, I-00146 Roma, Italy }

\author{A. Fondacaro}
\affiliation{European Synchrotron Radiation Facility, BP 220, F-38042 Grenoble
Cedex 9, France}

\author{S. Huotari}
\affiliation{European Synchrotron Radiation Facility, BP 220, F-38042 Grenoble
Cedex 9, France}

\author{P. Lacovig}
\affiliation{Sincrotrone Trieste S.C.p.A., Strada Statale 14 km 163.5,
I-34012 Trieste, Italy}

\author{O.~Tjernberg}
\affiliation{European Synchrotron Radiation Facility, BP 220, F-38042 Grenoble
Cedex 9, France}
\affiliation{Materials Physics, Royal Institute of Technology KTH, Electrum 229,
S-164 40 Kista, Sweden}

\author{N. B. Brookes}
\affiliation{European Synchrotron Radiation Facility, BP 220, F-38042 Grenoble
Cedex 9, France}

\author{M. Sacchi}
\affiliation{Synchrotron SOLEIL, Bo\^{\i}te Postale 48, F-91142 Gif-sur-Yvette,
France}
\affiliation{Laboratoire de Chimie Physique-Mati\`{e}re et Rayonnement,
Universit\'{e} Pierre et Marie Curie, UMR 7614, F-75005 Paris, France}

\author{P. Metcalf}
\affiliation{Department of Chemistry, Purdue University, West Lafayette, Indiana
47907, USA}

\author{G. Panaccione}
\affiliation{Laboratorio Nazionale TASC-INFM-CNR, in Area
Science Park, I-34012 Trieste, Italy}

\date{\today}

\begin{abstract}

We present a comprehensive study of the photon energy dependence of the 
valence band photoemission yield in the prototype Mott-Hubbard
oxide V$_2$O$_3$. The analysis of our experimental results, covering an extended photon energy range (20-6000 eV) and combined with GW calculations, allows us to identify the nature of the orbitals contributing to the total spectral weight at different binding energies, and in particular to locate the V 4$s$ at about 8 eV binding energy. From this comparative analysis we can conclude that the intensity of the quasiparticle photoemission peak, observed close to the Fermi level in the paramagnetic metallic phase upon increasing photon energy, does not have a significant correlation with the intensity variation of the O 2$p$ and V 3$d$ yield, thus confirming that bulk sensitivity is an essential requirement for the detection of this coherent low energy excitation.

\end{abstract}

\maketitle

The complex and fascinating physics of strongly correlated oxides has one of
the most remarkable examples in vanadium sesquioxide, V$_{2}$O$_{3}$, often
considered as a prototype Mott-Hubbard system, where the competition
between correlation and itinerant behavior of the electrons plays a crucial role in determining the electronic properties~\cite{mott,imada}. 
V$_{2}$O$_{3}$ presents a rich phase diagram and undergoes, as a function of temperature,
pressure and doping, a number of transitions passing from the paramagnetic
metallic (PM) phase to the antiferromagnetic insulating (AFI) one, as well
as from the PM to the paramagnetic insulator (PI) phase \cite{mcwhan}. 
In the former case, the metal-insulator transition (MIT) is accompanied by a
change of structure, from $\alpha -$corundum to monoclinic, whereas the
latter, obtained for instance by Cr doping, is isostructural. 

In the description provided by dynamical mean-field theory (DMFT) \cite{revmod},
the coexistence of coherent quasiparticle (QP) features close to the Fermi energy E$_{F}$ 
and incoherent lower Hubbard band (HB) at higher binding energy characterizes the correlated metallic phase of a Mott-Hubbard compound: V$_{2}$O$_{3}$ served as a test system to verify these predictions, and photoelectron spectroscopy (PES) played an important role to this end.  
In particular, high energy PES (using either soft or hard X-rays) combining the direct probe of the electronic density of states (DOS)
with enhanced bulk sensitivity \cite{sekiyama_00}, made it possible to reveal a clear coherent intensity near E$_{F}$ \cite{mo,mo2006,panaccione}. Such a pronounced structure couldn't be observed in previous attempts using lower energy photons and a consequently more surface sensitive detection.   
Recent LDA+DMFT calculations \cite{keller04,poteryaev,panaccione} provide a good agreement with the experimentally determined shape and intensity for this peak, 
making it possible to extract quantitative information on fundamental physical parameters for a correlated systems, such as the effective Coulomb interaction parameter U. 

Although a consensus has been reached for the presence of a QP intensity
close to E$_{F}$ in metallic V$_{2}$O$_{3}$, several questions still remain unanswered, and
call for a more detailed experimental and theoretical analysis of the V$_{2}$O$_{3}$
valence band, concerning in particular the role of the different V and O orbitals.  On the experimental side, the difficulty in revealing the QP
intensity has been mainly explained with the different electronic structure
between surface and bulk. In general, not only the surface preparation of V$%
_{2}$O$_{3}$ may substantially alter the structural and electronic
properties \cite{thomas,toledano}, but also significant differences in QP
intensity and lineshape could be found as a function of experimental
conditions like for instance the spot size of the probe \cite{mo2006}. This marked sensitivity to any perturbation of the bulk electronic structure has been recently interpreted as related to a surface dead layer where the QP vanishes, the essential physical reason being that the effects of the surface propagate over a characteristic length scale which is larger for strongly correlated materials  \cite{rodolakis2009,tosatti2009}. It is interesting to notice that the QP peak was found to give a prominent PES yield also when detected at very low (less than 10 eV) photon energy, where the level of bulk sensitivity of the photoelectrons increases again \cite{rodolakis2009}. In fact, all
these reports point out the importance of using bulk sensitive PES to obtain
reliable information of QP intensity and transfer of spectral weight at MIT.
At odd with this picture, cluster calculations on V$_{2}$O$_{3}$ suggest that the observed evolution of the QP spectral weight as a function of the photon energy should not be solely due to a change in surface vs bulk sensitivity, but that the change in V 3$d$/O 2$p$ cross section ratio upon changing photon energy may also play an important role, given the V 3$d$-O 2$p$ hybridization \cite{abbate1,abbate2}. Site specific PES experiments, together with density functional theory (DFT) calculations, 
investigated O-V hybridization: the theoretical analysis of the various contributions, together with fitted cross-sections and an approximate description of many body effects, could essentially reproduce the experimental spectrum taken at one specific photon energy (hv=2286 eV) \cite{woicik}.

The question of the interplay between hybridization and photon energy dependence of the spectra 
is hence one of the remaining key questions to be studied, in order to put the understanding 
of this complex material on a firmer basis. The prototype theoretical approach to the study 
of correlated materials, namely DMFT, suffers from the limitation that 
different kinds of orbitals are treated on different levels in order to keep the calculations feasible. 
Indeed, recent LDA+DMFT results for V$_2$O$_3$ \cite{poteryaev} obtain that both the QP and the satellite 
are dominated by $e_g^\pi$ states (while the $a_{1g}$ has only a minor contribution), but the fact that these calculations treat $p$ and $d$ orbitals on a different footing doesn't allow one to draw definite conclusions about the hybridization.   
Methods that treat all orbitals on the same footing are typically band structure approaches like DFT in the Kohn-Sham (KS) formulation 
or approaches that describe band structure and additional features due to dynamical correlation (satellites) 
for situations where the band structure picture is still dominant. 
The state-of-the art approaches of the latter kind is the GW approximation, 
where the self-energy is calculated as a product of the one-electron Green's function $G$ 
and the screened Coulomb interaction $W$ \cite{hedin}. 
Both the \textit{ab initio} KS and the GW approaches are free of empirical parameters, which is a crucial point for our goal: 
hybridizations should be predicted, and then tested {\it a posteriori}, not simply fitted to experiment. 
Moreover, it is known that quasiparticle features calculated in GW are reliable 
even when satellites are not very well described \cite{ferdigw}, 
so that the method can also be applied to transition metal oxides, 
as long as the question of interest is well defined in the quasiparticle framework.

We adopt hence in this work the KS and GW approaches for a theoretical description of the experimental spectra.
GW is needed in order to obtain reliable positions of all quasiparticle features 
(including the O 2$p$  and V 4$s$ dominated region). We use a quasiparticle self-consistent GW approach where wavefunctions and
 energies are improved with respect to KS \cite{faleev,fabien}. It turns out that, close to the findings for 
similar materials \cite{vo2cohsex}, the change of wavefunctions 
with respect to a KS calculation mostly concerns a mixing between $d$ states, 
whereas the $s$-$d$ and $p$-$d$ hybridization is already correctly predicted in KS-LDA. 
Concerning experiment, we present spectra consisting of extended valence band PES data from metallic stoichiometric V$_{2}$O$_{3}$ and Cr-doped (V$_{1-x}$Cr$_{x}$)$_{2}$O$_{3}$ (x= 0.11), covering a very large range of photon energies, namely from h$\nu$ = 19  to h$\nu$ = 5934 eV. A large set of data is thus available for what concerns both intensity and lineshape variation of QP, along with a measurement of the entire V 3$d$ and O 2$p$ valence band region. The joint interpretation of calculated and measured spectra yields a coherent picture if the evolution of the structures close to the Fermi energy with increasing photon energy is attributed to surface effects, whereas the hypothesis of dominant $p$-$d$ mixing would not be consistent.

High quality single crystals, were grown at Purdue University using a
skull melting technique \cite{harrison} and characterized by XRD and SQUID
measurements. PES experiments were performed using three different
experimental setups: the VOLPE\ spectrometer for Hard x-ray PES \cite
{torelli} (ID16 beamline, base vacuum 9x10$^{-10}$ mbar) and a Scienta
SES-2002 for soft X-ray PES (ID08 beamline, base vacuum 1x10$^{-10}$ mbar),
both located at the European Synchrotron Radiation Facility (ESRF), and a
Scienta SES-2002 for low energy PES (APE beamline, Elettra, Trieste - base
vacuum 1x10$^{-10}$ mbar). Spot size on the sample was 50 x
120 ${\mu}m^{2}$ and the overall energy resolution (beamline + analyser) was set to
450 meV (ID16 and ID08) and 50 meV (APE). The position of Fermi energy E$_{F}$ 
and the overall energy resolution were estimated by measuring the
metallic Fermi edge of polycrystalline Au foil in thermal and electric
contact with the samples. 
The specimens were carefully aligned and fractured in UHV to expose the (0001) plane, and the photoelectrons were detected at normal emission. All the data presented here were taken at 200 K, which for (V$_{1-x}$Cr$_{x}$)$_{2}$O$_{3}$ corresponds to the PM phase both for x=0 and x= 0.11, consistently giving identical results on several samples.

\begin{figure}
\includegraphics[width=\columnwidth]{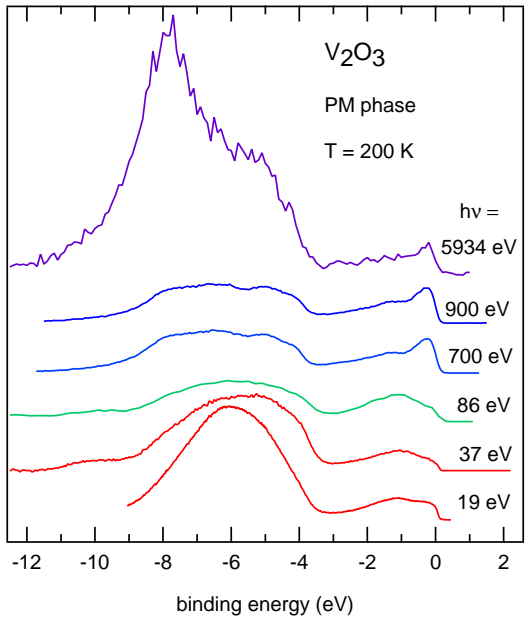}
\caption{(Color online)  
Experimental valence band PES spectra of V$_{2}$O$_{3}$, 
normalized to the bottom of the incoherent spectral
region at about -3 eV binding energy.}
\label{fig1}
\end{figure}

\begin{figure}
 \includegraphics[width=\columnwidth]{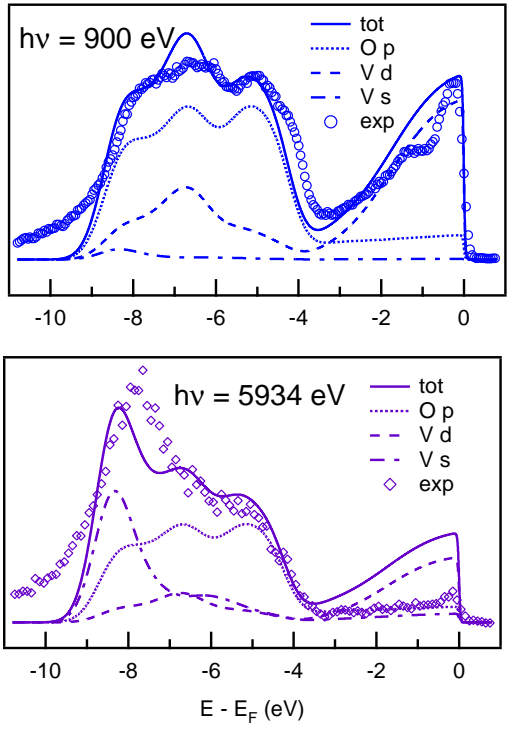}
\caption{(Color online) 
Measured valence band spectra compared to the total weighted DOS from GW calculations. The
weighting has been obtained by multiplying partial $s,p,d$ DOS's (see \protect Table~\ref{table1}).} 
 \label{fig2}
 \end{figure}

Figure \ref{fig1} collects the extended valence band PES experimental spectra vs. photon energy. In
each spectrum one recognizes two main spectral regions: one that we shall call VB,
usually defined as the O $2p$ valence region, from 3 to 11 eV, and the other
one between E$_{F}$ and 2-3 eV binding energy (BE), that we shall refer to as HB+QP.
As the photon energy increases in the low photon energy range (h$\nu$=19 eV to 86 eV), the intensity of the HB+QP increases, relative to the one of the VB region. In agreement with previous experimental results \cite{shinold}, spectra in this photon energy range present a pronounced and broad feature, roughly centred at 6 eV BE, evolving in a clear three peaks structure when soft x-ray (h$\nu$=700 eV-900 eV) are used (see Fig. \ref{fig2}), with a residual tail of intensity at BE 
$>$ 10 eV \cite{mo2006}. Interestingly, the spectral shape of the hard x-ray
PES (h$\nu $ = 5934 eV) spectrum is remarkably different, with a strongly
dominant VB region extending down to BE $\sim $12 eV, and a main peak at $\cong $ 8 eV BE. 
A similar peak structure has also been observed in Hard
X-ray valence band spectra of vanadates~\cite{eguchi}, cuprates \cite{PRBncco} and manganites \cite{offi}.

These results can be explained by the photon energy dependence of the cross sections 
of states with different symmetries as obtained in our calculations. In Fig. \ref{fig2}, indeed, the dotted, dot-dashed and dashed lines 
show the projection of the GW calculated density of states
\footnote{We have used the ABINIT code for the ground state and GW calculations \protect\cite{abinit}. Calculations are performed at the  experimental lattice parameters for pure V$_2$O$_3$ from \protect\cite{structv2o3}; we have checked that doping with Cr does not change significantly the DOS.  We have employed Troullier-Martins pseudopotentials \protect\cite{troull}, including V 3$s$3$p$ semicore states in valence. 
Convergence has been achieved with a 4x4x4 Monkhorst-Pack \protect\cite{monkpack} grid of $\mathbf{k}$ points, an energy cutoff of 90 Ha in LDA and of 45 Ha for the wave functions entering the self-energy. In quasiparticle self-consistent GW calculations, where a plasmon-pole model has been adopted, all the states in the energy range from -10 to +5 eV have been calculated self-consistently.}  
on O 2$p$, V 4$s$ and V 3$d$ contributions, respectively, multiplied by the Fermi distribution at 200 K. 
The contributions are then multiplied by a factor given by the corresponding change in cross section according to the used photon energy in the upper and lower panel, respectively, as taken from tabulated cross sections \cite{yeh,scofield} and shown in Table~\ref{table1} \footnote{For the spectrum at 5.934 KeV we have used the tabulated parameters at 6 KeV; for the spectrum at 900 eV we have taken an average of the values for 800 eV and 1000 eV.}. We apply a  Gaussian broadening of 0.4 eV simulating the experimental resolution. Moreover, since the incoherent part of the structure near the Fermi level is particularly strong as compared to the QP weight, in contrast to what one observes concerning the O 2$p$ group, we add a Lorentzian broadening $\eta(BE)=2.4-0.6BE$ for $BE < 4$ and thus simulate the redistribution of spectral weight that is not contained in our quasiparticle calculation. This procedure is justified if the HB and the QP are of same nature (i.e. V $d$ in the present 
 case), whereas in the case of strong $p$-$d$ hybridisation of the HB one should find a different result.

Although the only free parameter is the  different broadening that we use for the QP+HB and the O 2$p$ group, respectively, the good agreement with the experimental results is striking: not only is the qualitative 
behaviour nicely reproduced (e.g. the three-peak structure at 900 eV photon energy, 
evolving into a main peak with a low binding energy shoulder at 5934 eV), 
but also all absolute peak positions are in very good agreement with experimental structures 
\footnote{GW corrections shift the LDA peak positions in the VB region by -0.7 eV on average. Of course the tabulated cross sections are a rather rough estimate and results could be further improved by fitting, but this is not the scope of the present work.}.
This agreement allows us to draw unambiguous conclusions about the nature of the spectra. 
In particular, one can understand the strong peak that emerges at high photon energy at about 8 eV of binding energy as being entirely
due to vanadium 4$s$ states, and only to a minor extent to oxygen 2$p$.
The V 4$s$  photoionisation cross section becomes dominant at high photon energy: as it can be seen from the table, cross section 
of O 2$p$ and V 3$d$ states in the range 200-1400 eV are known to decrease 
in absolute value, while keeping almost constant their ratio \cite{mo2006,yeh}

\begin{figure}
\includegraphics[width=\columnwidth]{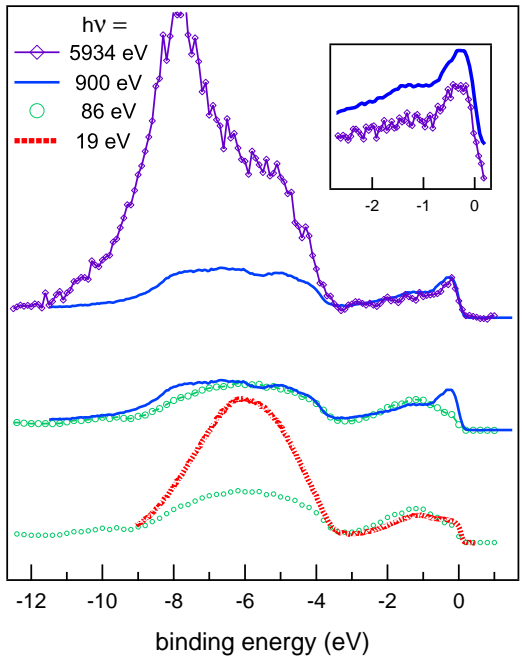}
\caption{(Color online) 
Direct comparison of selected pairs of experimental spectra taken at different photon energies. In the inset, the details of the QP region are shown for the two spectra at 900 and 5934 eV.}
\label{fig3}
\end{figure}

We now move to the inspection of the second region, HB+QP, by comparing
selected experimental spectra in Fig. \ref{fig3}. As was evident from Fig.~\ref{fig1}, a clear coherent intensity near E$_{F}$ is
observed at photon energies $\geq$ 700 eV. In this photon energy range the intensity ratio between the
QP region and the HB one is almost constant, despite the big variation of
intensity occurring in the VB region. In particular, the inset in Fig. \ref{fig3}
presents a zoom of the near E$_{F}$ region, from h$\nu =$ 900 eV and h$\nu$
= 5934 eV PES spectra. The width of the QP and relative spectral weight of
QP and HB are identical, within the noise level. Hence, the QP/HB ratio is not affected by the change of weight between V$s$ on one side and O$p$ and V$d$ on the other side. On the contrary, the comparison
between h$\nu$ = 900 eV and h$\nu$ = 86 eV spectra reveals that, although
the QP intensity vanishes in the spectrum at h$\nu$ = 86 eV, only minor
spectral weight changes are observed in the VB region. Consider that, at this photon energy, most of the intensity in the VB has an O 2$p$ character, due to cross section effects. Hence, if the change in the HB/QP ratio were dominated by O 2$p$ hybridisation, one should have also observed major changes in the VB. Finally, when h$\nu$ = 86 eV and h$\nu$ =
19 eV are compared, QP+HB region results to be similar, whereas the
intensity of VB region is highly enhanced at lower photon energy, which is again inconsistent with the hypothesis that O 2$p$ plays a strong role in the HB.

\begin{figure}
 \includegraphics[width=\columnwidth]{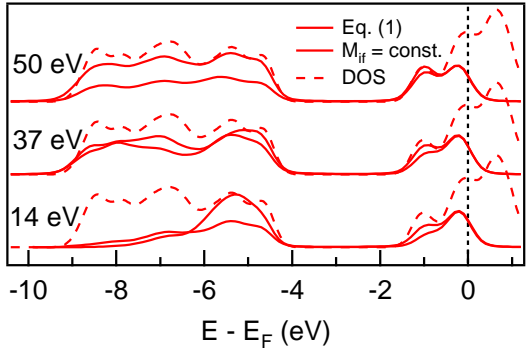}
 \caption{Calculated spectra for different photon energies (14, 37 and 50 eV) using Eq. \protect\ref{photocurrent}, 
including (continuous curves) or neglecting (dotted curves) matrix elements effects. The dashed curve (same for all photon energies) is the DOS.}
 \label{fig4}
 \end{figure}

The detailed variations within the low photon-energy range cannot be interpreted with tabulated cross sections; values are not available, and would moreover be hardly significant since at low energies, besides surface and many body effects, details of the density of empty states and material-specific matrix elements play a dominant role. In order to simulate these effects one should describe the photoelectron current $J_E$ by 
\begin{equation}
\label{photocurrent}
J_E(\omega) = \sum_{if} {\mid}M_{if}{\mid}^2 \delta(\epsilon_f-\epsilon_i-E)\delta(\epsilon_f-\omega). 
\end{equation}
In the one electron picture $M_{if}$ are dipole matrix elements between occupied and empty one electron states of energy 
$\epsilon_i $ and $\epsilon_f$, respectively, $\omega$ is the photoelectron energy and $E=h\nu$ is the photon energy. 
The common approximation to photoemission, namely the occupied DOS, 
is obtained when matrix elements $M_{if}$ are kept constant and the second $\delta$ function 
that represents energy conservation in the transition 
(i.e. a transition can only take place when final states in the reachable energetic region exist) is integrated over. In order to illustrate the qualitative importance of these contributions, we show in Fig. \ref{fig4} the calculated spectra corresponding to 14, 37 and 50 eV of photon energy, respectively \footnote{We have used the DP code \protect\cite{dpcode} with 216 shifted {\bf k}-points.}. The continuous curve corresponds to Eq. (\ref{photocurrent}), the dotted curve to the case where $M_{if}=const.$ is assumed. The dashed curve is the simple DOS, that does hence not vary with photon energy. Strong effects of both contributions are found in this energy range on  lineshape and intensity of the structures, and certain tendencies of the experiment can be detected (e.g. the fact that matrix elements tend to decrease the VB intensity with increasing photon energy). However, we stress the fact that  agreement with experiment on this detailed level might be in part fortuitous, since our calculations do neither include surface effects nor the partial angle resolution that is instead present in experiment. These graphs are instead meant to illustrate the sensitivity of low-photon-energy spectra to both effects contained in our calculations, and not expected to be cancelled by the additional experimental features. They imply that one should be careful to interpret spectra based on symmetry analysis and hybridizations only.

Finally, only within the 0-2 eV region, theoretical and experimental results differ.  
Whereas the analysis based on the decomposition of the QP on the atomic angular momenta
explains the main features in the VB region and their photon energy dependence,
it does not predict the strong depletion of the quasiparticle peak at low photon 
energies in the HB+QP region, and also our calculations containing the bulk final state and matrix elements do not explain the disappearance of the QP at low photon energy. 
Indeed, as already pointed out, while the VB region is dominated by one-particle states, 
the region close to E$_{F}$ is strongly influenced by incoherent structures due to strong correlation. Our results show that the measured spectra at  
h$\nu$ = 86 eV and h$\nu$ = 37 eV, while staying very similar in the 0-2 eV energy range, differ sizably in the VB region 
where O 2$p$  contribution is dominant as shown in Fig. \ref{fig3}). This 
gives clear evidence against the hypothesis \cite{abbate2} that the particular evolution of HB+QP can be due to the fact that the HB has a relevant O 2$p$ contribution while the QP is mainly V 3$d$. Our results, based on an equal-footing treatment of $p$ and $d$ orbitals, and on experiment, suggest that both QP and HB are strongly V $d$-dominated, and give a comprehensive picture that is
consistent with the hypothesis that QP and HB are influenced differently by the presence of the surface.

In conclusion, combining photoemission measurements performed at different photon energies with state-of-the-art GW calculations, we were able to disentangle the contribution of V 3$d$, V 4$s$ and O 2$p$ orbitals to the valence band photoelectron yield of the prototype Mott compound V$_2$O$_3$. Due to cross section effects, the V 4$s$ band becomes particularly prominent in the hard X-ray regime, which makes it possible to unambiguously determine its binding energy at 8 eV from E$_F$. The possibility of including the orbitals of both V and O in ab initio calculations represents an important step forward towards a complete understanding of this model system, and allowed us to extract novel information on issues that are of general interest for many multi-orbital strongly correlated materials, like the level of hybridization between different bands. For the specific case of V$_2$O$_3$ in its metallic phase, we were able to conclude that the QP peak and the lower HB have   essentially a V 3$d$ origin, and that the photon energy dependence of the QP/HB photoemission intensity cannot be explained only by O 2$p$/V 3$d$ hybridization and cross section effects. These results confirm that bulk sensitivity is an essential requirement for photoemission studies of the electronic structure of correlated materials.   

\begin{acknowledgments}
We acknowledge fruitful discussions with C. Giorgetti, P. Torelli, P. Wzietek and D. J\'{e}rome. This work was supported
by the 7th framework programme through the ETSF e-I3 infrastructure project (Grant agreement No. 211956), ANR project NT05-3 43900, the CEA nanoscience programme, and INFM-CNR. 
Work at Elettra was supported by the EEC through the initiative "Integrating Activity on Synchrotron and Free Electron Laser Science"'
\end{acknowledgments}

\bigskip \newpage

\clearpage

\begin{center}
\begin{table}
\caption{Calculated V 3$d$, V 4$s$ and O 2$p$ cross sections at
selected photon energies, from Ref. \protect\cite{yeh} (a) and Ref. \protect\cite{scofield} (b). The
values from Ref. \protect\cite{yeh}, labeled 1 keV and 8 keV for simplicity,
correspond to the tabulated values at 1041 eV and 8047.8 eV, respectively.}
\begin{tabular}{|c|c|c|c|c|c|c|c|}
\hline
\multicolumn{8}{|c|}{V and O cross sections at selected photon energies
(Mbarns/electron)} \\ \hline
                      & 21.2 eV               &   800 eV              & \multicolumn{2}{|c|}{1000 eV} 
                      & 6 keV                 &   \multicolumn{2}{|c|}{8 keV}               \\ 
\hline
V $3d$                &  $^{a}$1.94           &   $^{a}$1.3x10$^{-3}$ & $^{a}$0.5x10$^{-3}$ &
$^{b}$ 0.6x10$^{-3}$  &  $^{b}$0.45x10$^{-6}$ &   $^{a}$2x10$^{-7}  $ & $^{b}$1.3x10$^{-7}$ \\ 
\hline
V $4s$                &  $^{a}$0.09           &   $^{a}$9.5x10$^{-4}$ & $^{a}$0.5x10$^{-3}$ &
$^{b}$ 0.6x10$^{-3}$  &  $^{b}$1.2x10$^{-5}$  &   $^{a}$5.5x10$^{-6}$ & $^{b}$5.9x10$^{-6}$ \\ 
\hline
O $2p$                &  $^{a}$2.6            &   $^{a}$5.5x10$^{-4}$ & $^{a}$0.22x10$^{-3}$ & 
$^{b}$ 0.27x10$^{-3}$ &  $^{b}$3.1x10$^{-7}$  &   $^{a}$1x10$^{-7}$   & $^{b}$1x10$^{-7}$    \\ 
\hline
Ratio $3d/4s$         &  21.5                 &   1.37                & 1                    & 
1                     &  0.038                &   0.036               & 0.022                \\ 
\hline
Ratio $3d/2p$         &   0.75                &   2.36                &  2.27                &  
2.22                  &   1.45                &   2                   &  1.3                  \\ 
\hline 
\end{tabular}
\label{table1}
\end{table}
\end{center}

\clearpage

\clearpage



\clearpage

\clearpage

\clearpage

\clearpage

\end{document}